\documentclass[runningheads]{llncs}
\usepackage{graphicx}
\usepackage{epsfig}
\usepackage{calc}
\usepackage{amsmath}
\usepackage{booktabs}
\usepackage{mathpartir}		
\usepackage{xspace}			
\usepackage{enumerate}		
\usepackage{longtable}
\usepackage{color}			
\usepackage{ifthen}			
\usepackage[colorinlistoftodos]{todonotes}
\usepackage{color}

\usepackage{footnote}

\usepackage[normalem]{ulem}
\usepackage{comment}

\usepackage{array}
\newcommand{\PreserveBackslash}[1]{\let\temp=\\#1\let\\=\temp}
\newcolumntype{C}[1]{>{\PreserveBackslash\centering}p{#1}}
\newcolumntype{R}[1]{>{\PreserveBackslash\raggedleft}p{#1}}
\newcolumntype{L}[1]{>{\PreserveBackslash\raggedright}p{#1}}

\usepackage{subcaption}
\usepackage{listings}
\usepackage{url}

\usepackage[misc,geometry]{ifsym}

\begin{document}

\title{Regulatory compliance-readiness in the AI Supply Chain: examining datasets in Hugging Face}

\author{Georgia M. Kapitsaki\orcidID{0000-0003-3742-7123} \textsuperscript{(\Letter)} \protect}

\authorrunning{G. Kapitsaki}

\institute{University of Cyprus, Nicosia, Cyprus \\
\email{gkapi@ucy.ac.cy}
}

\titlerunning{Regulatory compliance-readiness in the AI Supply Chain}

\maketitle  

\begin{abstract}
A very large number of datasets are made available via Hugging Face (HF). These datasets are used to train a vast number of AI models, also available via HF. There are regulatory issues that might arise in the AI supply chain when those datasets do not follow, for instance, data privacy practices, or do not disclose their data sources, cleaning and preparation processes. In this preliminary work, regulatory compliance-readiness is examined as a quality attribute in the framework of HF datasets, with a main focus on data privacy (e.g. GDPR, CCPA). Towards this direction, an analysis on regulatory compliance (e.g. with privacy laws) of the dataset using the dataset card and the dataset structure overview as starting points has been performed. We collected 11,682 HF datasets that have a dataset card, are not gated and have at least 500 downloads and analyzed the datasets using automated techniques and manual analysis on a sample of the dataset. The results show that a very small number of datasets are explicit on the datasets creation processes and mention data sources, while some make mentions to Personally Identifiable Information (PII) or other sensitive data in their data schemas. These results show the need for more detailed examinations of regulatory compliance in AI datasets and models, and research on tools that provide guidance to practitioners and automate the compliance process.

\end{abstract}

\section{Introduction}

Hugging Face (HF) is one of the most popular platforms for hosting AI models and datasets~\cite{jain2022hugging,kadasi2025model}.
A very large number of datasets are made available, with the datasets being used to train a vast number of AI models, also available via HF. In early 2026, HF hosts almost 3 million models and close to 1 million datasets. Regulatory compliance of AI datasets is becoming vital nowadays and regulations are handling how data are treated in datasets and how they are used for training purposes. For instance, provisions of the EU AI Act are being gradually enforced with a timeline covering 2024 to 2027, e.g. prohibited practices like AI exploiting vulnerabilities, and requirements for AI literacy have been applied in early 2025~\cite{act2024eu}. At the same time, data protection regulations, such as the EU General Data Protection Regulation (GDPR)~\cite{voigt2017eu} and the California Consumer Privacy Act (CCPA)~\cite{bonta2022california}, are highly relevant as they refer to users' consent for data collection and processing purposes, data portability and data minimization. GDPR also includes the right to avoid automated decision among the main eight user rights (along with right to be informed, right of access, right to rectification, right to erasure, right to restrict processing, right to data portability and right to object) that becomes more timely relevant with the rise of AI automation processes for data analysis.

Prior work has analyzed the parameters reported in HF models, including licensing information and bias~\cite{longpre2024large,pepe2024hugging}, whereas a similar analysis has been performed for HF datasets including license compliance~\cite{stalnakerempirical}. Other parameters concerning the use of the models have also been examined, such as in creating a graph showing links between models and datasets~\cite{rahman2025hugginggraph}. Nevertheless, no prior work has addressed the extent to which dataset cards are compliant-ready when it comes to data sources and preparation, as well as the potential inclusion of sensitive data. This is an important area of study in order to understand the current AI datasets landscape and the implications that might come from an even wider use of the existing datasets in AI models or even for other purposes.

Taking into consideration the above, in this work we have performed a preliminary analysis on the completeness and regulatory compliance readiness (e.g. for data privacy law compliance) linked with HF datasets, as an important quality attribute in the AI supply chain. The ultimate goal of the work is to provide a new line of research where regulatory compliance issues are examined more thoroughly in AI datasets and models. We collected 11,682 HF datasets that are not gated and have at least 500 downloads, and have relied on the respective dataset cards and data schemas overview for the analysis. We examined the collected datasets using automated analysis on specific sections and mentions encountered in the dataset cards (e.g. sources of data), while we partially used manual verification on a sample of the dataset.

With a focus on the above main aim, the following Research Questions (RQs) that include different specific sub-questions are introduced:
\begin{itemize}
    \item \textbf{RQ1. Do the dataset cards indicate the data sources, collection, preparation and processing activities performed?} Are the dataset cards explicit on the origin of data and their categories, as well as data cleaning processes? To which extent are datasets using anonymization or other privacy presentation techniques? This set of questions aims to examine transparency of data collection and preparation processes in the dataset cards.
    \item \textbf{RQ2. Do the datasets mention specific laws and do they include sensitive information in their data schemas?} Do the datasets on HF indicate data privacy laws or other regulations (e.g. EU AI Act)? Do the datasets include personal or sensitive information in their data schemas? Do the cards indicate dataset risks? These questions aim mainly to find whether Personally Identifiable Information (PII) or other sensitive data are found in the column names of the datasets.
\end{itemize}

The results show that the vast majority of datasets are not being explicit on the datasets risks and do not mention data sources, while some keep mentions to PII or sensitive information in their data schemas, and have not used privacy preservation techniques. Via analysis on a sample of the datasets (162 dataset cards), a very small number of datasets that contain user-identifiable information, such as usernames, were found. 

\textbf{Data availability}. The dataset cards of the analyzed datasets, the main data analyzed and the analysis scripts are made available for replication purposes on FigShare: \url{https://figshare.com/s/5b6c87995718d81096fd}.

The remainder of the text is structured as follows. Section 2 describes the regulatory landscape and related work in the area, while section 3 focuses on the methodological approach presenting the analysis areas and the data collection and analysis processes. Results are presented for each RQ in section 4 and are further discussed in section 5. Finally, section 6 concludes the work.

\section{Background and related work}

\subsection{Regulatory landscape}

As we are focusing on privacy leaks and relevant regulatory issues concerning the data preparation, we are considering mainly data privacy laws, such as the General Data Protection Regulation and the California Consumer Privacy Act~\cite{bonta2022california,voigt2017eu}. An overview of how main laws are relevant to the context of the current work is provided in the following:

\begin{itemize}
    \item \textbf{Core data privacy laws}: The EU GDPR, CCPA and its amendment California Privacy Rights Act (CPRA)~\cite{determann2020california}, cover core user rights, such as the right to erasure and right to access, and are relevant as user data may form part of datasets used in HF models. Moreover, Brazil's Lei Geral de Proteção de Dados (LGPD) is strongly linked with AI models and datasets, as it includes specific rights for data subjects to request the deletion of their data from a model's training set. 
    \item \textbf{More recent privacy laws}: Amended Personal Information Protection Act (PIPA) of South Korea focuses on transparency and safety requirements for datasets used in Foundation Models~\cite{kim2024automated}. PIPA allows for the use of publicly available data (scraped from the web) for AI training without explicit consent. Moreover, China's Personal Information Protection Law (PIPL) is very strict on automated decision making and cross-border data transfer. PIPL indicates moreover unique requirements for the special case of image/face data scraping. India's Digital Personal Data Protection (DPDP) Act puts a lot of emphasis on user consent, as it indicates that it is required to clearly define the purpose of data processing.
    \item \textbf{More specialized privacy laws}: Biometric Information Privacy Act (BIPA) of Illinois (United States) is relevant but is a more specialized case, as it specifically targets Biometric data (faces in image datasets).
    \item \textbf{AI laws}: The EU AI Act mandates Data Governance (Article 10), requiring training data to be relevant, representative, and to the best extent possible, free of errors~\cite{act2024eu}. Some of the law parts have not been enforced yet (some expected for 2027), therefore readiness might be more limited. GDPR is also relevant here, since when a dataset contains personal data of EU residents, a valid legal basis (e.g. consent or legitimate interest) under GDPR must exist in order to use it for training.
\end{itemize}

\subsection{Analyzing Hugging Face}

Models and datasets from Hugging Face have been analyzed in prior works, with the vast majority of works focusing on models. Information on a large fraction of Hugging Face model cards and a respective dataset is provided in~\cite{suryani2025model}, while another work focused on the synchronization between HF and GitHub models~\cite{ajibode2026synchronization}. Naming practices were also analyzed via the releases of 52,227 Pre-trained Language Models (PTLMs) on HF and it was found that 148 different naming practices are used in PTLM releases~\cite{ajibode2025towards}. The automatic classification of pre-trained models for Software Engineering tasks was enabled in a prior work using data from HF~\cite{di2024automated}. A public dump of HF was used for this purpose. On a similar line of work, a taxonomy consisting of 147 SE tasks was created using HF models~\cite{gonzalez2025cataloguing}.

It was also documented how ethical considerations are reported in models from Hugging Face and GitHub~\cite{gao2024documenting}. The authors of this work analyzed manually 265 documents. Six main ethical themes were used: model behavioral risks (e.g. bias, AI hallucinations), model use cases (e.g. malicious use cases), model risk mitigation: (e.g. how to safety filter the model), data quality concerns (e.g. private data or biased labeling) that is also relevant to the current work, reference to other materials (links to external ethics audits), and others (general safety disclaimers). A similar work analyzed the parameters reported in HF models, including licensing information and bias~\cite{pepe2024hugging}. This work also analyzed whether and how pre-trained transformer models document the datasets they have been trained on.

A graph with links between models and datasets is provided in~\cite{rahman2025hugginggraph}. The graph transformed models and datasets into a structured graph (containing over 400,000 nodes and 460,000 edges). The authors used Natural Language Processing and Named Entity Recognition to scan dataset and model cards for hidden mentions of dependencies.

An extensive analysis covering both models and datasets with 760,460 models and 175,000 datasets extracted from Hugging Face was performed, in order to evaluate the current state of documentation~\cite{stalnakerempirical}. Real-world examples of shortcomings and suggestions for improvement were provided. A set of tools and standards to trace data lineage were explored in~\cite{longpre2024large}. The resulting Data Provenance Explorer tool  allows developers to trace the origin, lineage and legal permissions of popular AI datasets.

\subsection{Regulatory compliance}
    
When it comes to data privacy laws, prior work has addressed compliance in software systems, such as how it is handled during software development in open-source software repositories from GitHub~\cite{hennig2026whos,kapitsaki2025evolution}. Other works focused on Software Bill of Materials (SBOM) that provides a list of the dependencies of a software application. An empirical study investigated its adoption on GitHub~\cite{nocera2023software,nocera2025adoption}. 186 repositories that use Software Package Data Exchange (SPDX) and CycloneDX were identified, and it was concluded that the adoption of SBOMs is low. The adoption of SBOMs was also examined via the analysis of 4,786 GitHub discussions across 510 SBOM-related projects~\cite{bi2024way}. Moreover, a dataset with over 78 thousand unique SBOM files, deduplicated from files coming from over 94 million repositories was made available~\cite{soeiro2025wild}. 

In AI, a similar notion exists in AI Bill of Materials (AIBOM) and the emerging new version of SPDX, SPDX 3.0, has a dedicated AI Profile~\cite{bennet2025implementing}. For a more specific case, the Data Bill of Materials for AI (AI DataBOM) that needs to contain details about consent information and licenses of copyrighted works was previously introduced~\cite{zhang2024protecting}. 
 ALOHA is a tool that automatically generates CycloneDX-compliant AIBOMs for existing Hugging Face models~\cite{d2025aloha}. Also the AIBOM Generator\footnote{\url{https://genai.owasp.org/owasp-aibom/}} is available by OWASP (Open Worldwide Application Security Project).

In the area of law enforcement, a user study with the participation of criminal investigators was performed in order to learn about current practices, understand perspectives on AI-assisted systems and defining specific needs~\cite{nowack2025towards}. Participants verified the need of tools that analyze data and make predictions, with humans involved in the process of decision making. In the same line of work, the Responsible AI Monitoring Platform (RAMP) is a human-centered
system that supports experts in making AI governance
systematic and transparent~\cite{zielinski2026operationalizing}.

Privacy risks in code models were examined in~\cite{yang2025understanding}. It was examined more specifically whether the training dynamics of LLMs used for code generation on each type influence its leakage risk at inference time, and whether this relationship is causal. This work relied mainly on PII identification using source code files. The work closest to the current work is in the areas of biometrics and healthcare where over 100 datasets were surveyed and 60 datasets were analyzed~\cite{mittal2024responsible}. Via this process it was found that a susceptibility to fairness, privacy and regulatory compliance issues exists. Concerning privacy, some cases of privacy leakage were found with those cases referring to names, sensitive and protected attributes, accessories, critical objects, location inference and medical conditions.

\textbf{Relation to prior works}. While prior works have mostly focused on HF models~\cite{suryani2025model}, we shift the focus to how datasets used in these models are rich in their descriptions and ready for regulatory compliance, and we identify areas that need improvement
. Prior works examined provenance and licensing reporting practices in HF~\cite{gao2024documenting,rahman2025hugginggraph}. We focus instead on other areas of the dataset cards and schemas (e.g. ``\emph{Source Data}", ``\emph{Initial Data Collection}" sections). The closest work relies on a small number of datasets from specific domains, biometrics and healthcare, to find privacy leakage cases~\cite{mittal2024responsible}.

\section{Methodological approach}

\subsection{Analysis areas}

Considering the requirements that stem from the laws discussed in the previous section, the analysis is focused on the following main items related to the introduced RQs:
\begin{itemize}
    \item Relevant to RQ1: 
    \begin{itemize}
        \item Does the dataset card provide information on the origin of the data (data sources), their categories, how they were labeled/cleaned, and the assumptions made during preparation for regulatory audits? This is relevant to the EU AI Act and also data privacy laws, such as PIPA that focus on transparency.
        \item Do the dataset cards make mention of anonymization, pseudonymization, or differential privacy techniques applied? This is relevant to GDPR and other data privacy laws, and to the EU AI Act.
    \end{itemize}
    \item Relevant to RQ2: 
    \begin{itemize}
        \item Does the dataset schema preview contain any PII or other sensitive terms (e.g. email)? This is relevant to GDPR and other data privacy laws.
        \item Does the metadata of the dataset card prove that the dataset does not include privacy risks? This is relevant to the EU AI Act (Article 10 on Data Governance)~\cite{niemiec2026will}. For this part, we checked whether the dataset card contains a ``\emph{Bias, Risks, and Limitations}" section but we did not focus on further analysis of bias (that is the main purpose of the section), as this has been addressed in prior work~\cite{gao2024documenting,pepe2024hugging}.
    \end{itemize}
\end{itemize}

Since licenses of datasets have been examined in a recent work, we have not included this aspect in our analysis, although it is relevant to regulatory compliance~\cite{stalnakerempirical}.

\subsection{Data collection}

As aforementioned, we selected to use in this work the top datasets from the HF platform based on the number of downloads. HF was selected as in prior work, it was found that HF offers diverse features and data that make it a good source of empirical studies concerning the development, evolution and usage of AI-related projects~\cite{ait2025suitability}. HF has also been widely used in prior empirical works~\cite{gao2024documenting,gonzalez2025cataloguing,rahman2025hugginggraph}. 

We used the HF Endpoints API to collect the dataset card and schema of datasets that were among the top downloaded, and kept the ones having model card documents~\cite{hfendpoints}. Data collection took place in April 2026 (between April 17th and April 28th). Specifically, information from datasets that have at least 500 downloads was collected. Using this threshold ensures the analysis focuses on datasets with established community trust and high utilization, making them representative of the most influential and widely deployed models in the Hugging Face ecosystem. This number accounts for 1.75\% of all datasets hosted on HF according to the total number indicated in the platform for datasets in April 2026 (16,996 out of the 972,413 datasets). In HF all datasets with a card include a README.md file and we relied on this information in order to collect only datasets with a dataset card. Using this approach we collected the 12,293 most downloaded dataset cards and schemas. Prior work also offers datasets of dataset cards but some works are older (collected in July 2024~\cite{stalnakerempirical}), or focus on model cards instead of dataset cards~\cite{pepe2024hugging,suryani2025model}. For this reason, the HF dataset cards were collected  via the respective Endpoints API. Even though this resulted in a smaller number of dataset cards in comparison to using existing datasets, we were able to examine the currently most downloaded datasets as explained above, and collect also the respective data schemas offered by the dataset viewer of HF (not available in datasets from earlier works). 

611 (4.97\%) datasets (with at least 500 downloads and a dataset card) are gated and they were also left out from the analysis, as they require accepting specific terms for their use. That would hinder making the dataset used in this work publicly available, as they may impose specific redistribution rules or rules for analysis. Nevertheless, this is an indicator that a dataset might include sensitive data and the relevant percentage is non negligible. Keeping only datasets that have a (non-empty) dataset card and are not gated, we were left with 11,682 datasets. 

\subsection{Data analysis techniques}

In order to detect potential non-compliant ready cases in the datasets and their cards, we relied on both automated and manual techniques. For automated techniques, we relied on the structure of dataset cards on HF, where, for instance, the ``\emph{Bias, Risks, and Limitations}" section is a formal component of the HF metadata standard, although its presence is not strictly enforced by HF. Prior work has also found that this section is used in a limited way~\cite{yang2023navigating}. We thus, relied on relevant dataset card sections, and other attributes and endpoints from the HF API (listed in Table~\ref{tab:attrs-sections-examined}). We used simple statistics for this analysis. 

\begin{table}[!th]
\vspace{0.25cm}
\caption{Dataset card sections, attributes and endpoints utilized.}
\label{tab:attrs-sections-examined}
\centering
    \begin{tabular}{p{3.6cm}p{8cm}}
    \toprule
\textbf{Section} & \textbf{Description}\\
\midrule
Source Data & Describes the origins of data, indicating data collection and selection, source data producers, as well as validation and cleaning.\\
\midrule
Annotations &  Describes the annotation process, e.g. tools used, validation processes and annotators.\\
\midrule
Initial Data Collection, Cleaning, Preprocessing & These sections cover dataset creation processes.\\
\midrule
Bias, Risks, and Limitations & Covers the specific areas to indicate where the data might fail or cause harm, covering also mentions for personal and sensitive information.\\
\midrule
    \textbf{Attribute} & \textbf{Description} \\
    \midrule
gated	&Indicates whether the dataset is gated, i.e. requires users to request access and agree to specific terms before they can download/view data.\\
\midrule
task\_categories & HF categorizes tasks into 6 main domains.\\
\midrule
    \textbf{Endpoint} & \textbf{Description} \\
    \midrule
\texttt{info} & Returns the schema that lists the column names from the dataset.\\
\bottomrule
\end{tabular}
\vspace{-0.2cm}
\end{table}

We also created a list of relevant keywords to search for in the cards and the datasets schema. The list created covers a wide range of PII and sensitive data. We relied on existing works that refer to Personally Identifiable Information, such as username, email, password, name, IP address~\cite{yang2025understanding}, and to sensitive data~\cite{rumbold2018data}. The categories and the respective keywords in each category that we used are listed in Table~\ref{tab:privacy-keywords}. We used term frequency (TF) analysis for the keywords presence, and for the presence of other terms in the dataset cards sections. For the partial manual analysis, a representative sample of dataset cards was inspected by a human expert. 

\begin{table}[!th]
\vspace{0.25cm}
\caption{Keywords used to detect PII and other sensitive data.}
\label{tab:privacy-keywords}
\centering
    \begin{tabular}{lp{9.6cm}}
    \toprule
    \textbf{Category} & \textbf{Keywords} \\
    \midrule
    Identification & SSN, SOCIAL\_SECURITY\_NUMBER, PASSPORT, DRIVER\_LICENSE, VIN (Vehicle ID), IP\_ADDRESS, MAC\_ADDRESS\\
    \midrule
    Contact & EMAIL, PHONE\_NUMBER, STREET\_ADDRESS, ZIP\_CODE, USER\_ID, USERNAME, AUTHOR \\
    \midrule
    Personal & PERSON, NAME\_STUDENT, FIRST\_NAME, LAST\_NAME, GENDER, BIRTH\_DATE, AGE \\
    \midrule
    Medical/Genetic & PATIENT\_ID, MEDICAL\_RECORD\_NUMBER, DOCTOR\_NAME, HOSPITAL\_NAME, DIAGNOSIS, TREATMENT\_PLAN, LAB\_RESULT, BLOOD\_TYPE, MEDICATION, SURGERY\_DATE, GENE\_SEQUENCE, BIOMETRIC\_DATA, DNA\_MARKER \\
    \midrule
    Banking & CREDIT\_CARD, CVV, ACCOUNT\_NUMBER, ROUTING\_NUMBER, IBAN, SWIFT\_CODE, BITCOIN\_ADDRESS \\
    \midrule
    Employment & COMPANY\_NAME, JOB\_TITLE, SALARY, EMPLOYEE\_ID, TAX\_ID, ITIN \\
\bottomrule
\end{tabular}
\vspace{-0.2cm}
\end{table}

\section{Results}

\subsection{Dataset overview} 

The datasets analyzed have an average size of 454,973.5 MB (ranging from 0.787 kB to 306.85 TB), and are used on average in 6.53 AI models on HF (ranging from 0 to 1,435 models). There are 4 datasets that are used in more than 1,000 models (\emph{HuggingFaceH4/ultrafeedback\_binarized}, \emph{teknium/OpenHermes-2.5}, \emph{Hugging FaceH4/ultrachat\_200k}, \emph{microsoft/orca-math-word-problems-200k}). The 20 most popular tasks the datasets are used for are listed in Table~\ref{tab:dataset-tasks}, while there are in total 98 different tasks listed in the dataset cards. Most tasks are related as expected with text followed by images, while some datasets are also linked with audio and video. 

\begin{table}[!th]
\vspace{0.25cm}
\caption{Main task categories of datasets analyzed.}
\label{tab:dataset-tasks}
\centering
    \begin{tabular}{rlr}
    \toprule
    & \textbf{Task category} & \textbf{Frequency} \\
    \midrule
1 & text-generation	&1,118 \\
2 &question-answering	&817\\
3 &text-classification	&585\\
4 &robotics	&524 \\
5 &image-classification	&406 \\
6 &visual-question-answering&	312 \\
7 &image-to-text	&259\\
8 &text-retrieval	&239\\
9 &object-detection&	234\\
10 &automatic-speech-recognition	&217 \\
11 &text-to-image	&216\\
12 &image-segmentation&	202\\
13 &feature-extraction	&188\\
14 & audio-classification	&164\\ 
15 & image-to-image	&162\\
16 &text2text-generation	&141\\
17 &summarization	&138\\
18 &image-text-to-text&	135\\
19 &text-to-speech	&135\\
20 &translation	&125\\
\bottomrule
\end{tabular}
\vspace{-0.2cm}
\end{table}

\subsection{RQ1. Do the dataset cards indicate the data sources, collection,
preparation and processing activities performed?}

In this RQ, we examined whether the dataset card indicates the data sources used, the collection process and any processes applied for privacy preservation purposes, such as anonymization, pseudonymization, or differential privacy techniques. 

Concerning data sources (by looking for the presence of a ``\emph{Source Data}" section), there are 3.47\% of datasets (405 datasets) with such a mention.  We had a more thorough look at the top 50 datasets (in terms of number of downloads) that are specific on their sources and the following sources were found among the descriptions of the dataset card: CommonCrawl (e.g. WARC, WAT dumps), ArXiv, data created from scratch, Protein Data Bank, ACL Anthology, Semantic Scholar, instructables.com, parliamentary proceedings, Wikipedia, archives.gov, Reddit dumps, GitHub repositories, ClickHouse Playground, Yahoo Finance, conference competitions, European Parliament event recordings, Epstein files, AmbientCG, CGBookCase, PolyHaven, ShateTexture, TextureCan, elementary/ high school science curricula, Yahoo Flickr dataset, archive.org, BBC News, images. Many cases had references to other HF datasets for more information (9 out of 50), and also the ones that did not make a reference to a specific data source were referring to other HF datasets. In the whole dataset with this section, ArXiv was the most frequently encountered data source (31 cases), followed by CommonCrawl and Wikipedia (each 26 cases).

The presence of ``\emph{Annotations}" section is scarce. 356 (3.05\%) of datasets have such a section and 337 (2.88\%) have a text other than mentioning that no annotations are contained. In this section, there are many references to external research publications and GitHub repositories. Concerning cleaning processes on the data (``\emph{Initial Data Collection}", ``\emph{Cleaning}", ``\emph{Preprocessing}" sections), the vast majority of dataset cards do not indicate such a section, and the few that do (323 or 2.76\% of datasets) refer primarily to specific sections in respective publications (e.g. \emph{openai/gsm8k} dataset). 

Although the dataset cards do not include generally the main sections for data collection and processing, a larger percentage uses privacy preservation techniques: 11.21\% of datasets (1,309 datasets) mention privacy preservation techniques, with the techniques indicated listed in Table~\ref{tab:privacypreserv-freqs}. Most datasets rely on synthetic data or use anonymization and aggregation processes. Concerning PII, HF users might use specific scrubbing techniques with models like PII-BERT\footnote{\url{https://huggingface.co/ab-ai/pii_model}} or NVIDIA's Nemotron-PII\footnote{\url{https://huggingface.co/datasets/nvidia/Nemotron-PII} datasets} to scan datasets before uploading so we have also searched specifically for these cases~\cite{haleluyah2025towards}. We found 20 (0.19\%) datasets with an indication of scrubbing: Presidio (4 cases), Nemotron-PII (2 cases), de-identified (15 cases), general mention to scrubbing (2 cases), so overall the use of such techniques is rare, or is at least not documented in the dataset card.

\begin{table}[!th]
\vspace{0.25cm}
\caption{Privacy preservation techniques in datasets.}
\label{tab:privacypreserv-freqs}
\centering
    \begin{tabular}{lrrr}
    \toprule
    \textbf{Technique} & \textbf{Frequency} & \textbf{Percentage} & \textbf{Percentage}\\
    && \textbf{(all datasets)} & \textbf{(with techniques)}\\
        \midrule
Synthetic data & 532 & 4.55\% & 40.64\%\\
Anonymization	&246 & 2.11\% & 18.79\%\\
Aggregation	&238 & 2.04	\% & 18.18\%\\
Pseudonymization&	213 & 1.82\% & 16.27\%\\
Differential privacy&	130 & 1.11\% & 9.93\%\\
Redaction	&94 & 0.8\% & 7.18\%\\
Access restrictions	&24 & 0.21\% & 1.83\%\\
\bottomrule
\end{tabular}
\vspace{-0.2cm}
\end{table}

By taking a look on datasets that use more than one from the above sections and mechanisms in dataset card descriptions, we found only two cases that follows all apart from scrubbing (\emph{jmhessel/newyorker\_caption\_contest} and \emph{creative-graphic-design/GenPoster100K} datasets). In one of them, the sections only only refer the reader to the respective publication  (\emph{jmhessel/newyorker\_caption\_contest} dataset). There are also only 11 datasets that contain all sections but do not indicate privacy preservation or scrubbing. 

\subsection{RQ2. Do the datasets mention specific laws and do they include
sensitive information in their data schemas?}

In this RQ, we searched for mentions of PII and other sensitive data within the dataset schemas, as well as for mentions to main relevant laws. It was not feasible to obtain column names for a large number of datasets: 3,952 (33.83\%) datasets, so the results reported are based on the 7,730 datasets with column names that could be retrieved from HF. This is attributed to the use of formats other than CSV, JSON Lines, Parquet or Txt for describing the data schema, and hence HF API cannot export the schema.

For the laws, we relied on the main laws introduced earlier in the text (GDPR, CCPA, CPRA, AI Act, PIPL, LGPD, PIP, BIPA, and in addition Health Insurance Portability and Accountability Act - HIPAA) and used for search purposes the full name of the law and its abbreviation. Mentions to specific laws are very scarce within the dataset cards. Only 30 datasets make such a mention with GDPR being the most frequent case (Table~\ref{tab:law-freqs}). Thesse mention refer to the dataset being compliant to the law, e.g. \emph{PleIAs/common\_corpus} dataset mentions both GDPR and the AI Act for this purpose, while \emph{ontocord/VALID} refers to the AI Act in its ethical considerations to showcase the transparency and safety elements of the dataset.

\begin{table}[!th]
\vspace{0.25cm}
\caption{Law mentions in dataset cards.}
\label{tab:law-freqs}
\centering
    \begin{tabular}{lr}
    \toprule
    \textbf{Law} & \textbf{Frequency} \\
        \midrule
    General Data Protection Regulation	&22\\
    AI Act &9\\
    Health Insurance Portability and Accountability Act &7\\
    California Consumer Privacy Act &5\\
\bottomrule
\end{tabular}
\vspace{-0.2cm}
\end{table}

We calculated also the frequencies of the most common PII and other sensitive data that can be found in the column names using the terms indicated in the \emph{Data analysis techniques} section of the current work, with 277 (2.37\%) datasets containing at least one PII or other sensitive data mentions. The frequencies covering both categories are listed in Table~\ref{tab:pii-sensitive-freqs}, with most cases having a reference to key or name (we excluded cases that were referring for instance to repository names or filenames). Concerning the use of scrubbing and privacy preservation techniques that were targeted in RQ1, the datasets that contain scrubbing do not contain any PII or other sensitive data, while 47 cases with PII that use privacy preservation techniques (refer to RQ1 analysis) mention that they used anonymization (18 cases), synthetic data (18 cases), aggregation (12 cases), differential privacy (4 cases), pseudonymization (3 cases), redaction (2 cases) and access restrictions (1 case).

\begin{table}[!th]
\vspace{0.25cm}
\caption{PII and sensitive data indications in column names.}
\label{tab:pii-sensitive-freqs}
\centering
    \begin{tabular}{lrr}
    \toprule
    \textbf{PII/Sensitive keyword} & \textbf{Frequency} & \textbf{Percentage (with schema)} \\
        \midrule
author&	121 & 1.57\%\\
gender&	108 & 1.4\%\\
age	&69 & 0.89\%\\
user/username&	14 & 0.18\%\\
email &	5 & 0.06\%\\
diagnosis	&5 & 0.06\%\\
person	&4 & 0.05\%\\
vin	&3 & 0.04\%\\
salary	&2 & 0.03\%\\
\bottomrule
\end{tabular}
\vspace{-0.2cm}
\end{table}  

For a more fine grained analysis, we focused on a sample of datasets that contain PII/sensitive terms. The aim of this manual analysis was to verify whether sensitive data are included, since using the column names is not a measure that can be used alone to detect sensitive data. We chose a statistically significant sample of projects that was randomly chosen (using the \texttt{sample} function in R) with a confidence level of 95\% and a margin of error of ±5\% (using Cochran’s formula). 162 datasets were thus, analyzed. This manual analysis process was performed by the author (senior researcher with expertise on empirical software engineering and GDPR compliance). As the task was straightforward no second coder was used. For the manual analysis the dataset card was inspected online. On average, 2 minutes were spent on each dataset. 

After the manual inspection of the datasets, we calculated how many cases refer indeed to sensitive data. We used precision for this purpose: $precision=\frac{real\_sensitive\_retrieved}{all\_sensitive\_retrieved}=0.51$~\cite{buckland1994relationship}. This can be used as an indication that among the cases detected via the keyword-based search, the real cases where sensitive data were not handled appropriately are 142 (1.84\% of the datasets with a schema returned by HF). In the sample, some examples of cases with false positives were referring to using `author' as column name for a book author (e.g. \emph{howard-hou/OCR-VQA} dataset) or whether the author is human (e.g. \emph{toxigen/toxigen-data} dataset) or hashing the actual value of the column (e.g. \emph{Exorde/exorde-social-media-december-2024-week1} dataset). In other cases, data present in the column names were actually encoded in the dataset or were empty (e.g. \emph{amu-cai/C3T} dataset). Among the cases with real sensitive data in the datasets, there were 20 datasets (12.25\% of datasets in the sample) where usernames from various locations were present (e.g. GitHub, Reddit). We found also a problematic case that includes images of persons, while not indicating details about how the images were retrieved, and one case that indicates that images were obtained from an open dataset (\emph{Harvard-Edge/Wake-Vision}).

Finally, concerning the presence of a ``\emph{Bias, Risks, and Limitations}" section that may indicate sensitive data included in the dataset, 6.81\% of the dataset cards (796 datasets) contain such a section. We inspected manually the top 50 datasets with such a section in order to examine whether privacy risks are mentioned. We found only three mentions of either not using any private data or not being able to control private data (\emph{Divotion/SR-Ground}, \emph{Skywork/SkyPile-150B}, \emph{Flori83/TroveLedger} datasets). We also encountered references to legal issues in datasets by \emph{mlfoundations}, \emph{HuggingFaceFW} and \emph{Salesforce} accounts.

\section{Discussion}

Our results show overall that specific mentions to data risks, as well as techniques to minimize privacy concerns (e.g. anonymization) are scarce in the dataset cards. This is not a direct indicator that a dataset carries such risks, but it is an indicator that the dataset creators are not paying special attention to including these important details in the dataset descriptions. Many datasets cards with these sections provide external references (mainly to research publications/pre-prints). So overall, the information does not exist in HF or is scattered in different locations, outside HF. Concerning sensitive data, although many false positives were found in relevance to the keywords-based search, the percentage of datasets that contain sensitive data of high risks (12.25\% in the sample analysis) is high and indicates that many datasets do not take measures for privacy-preservation (e.g. to remove usernames when these are not needed, or when the users have not given prior consent).

\textbf{Threats to validity.} Our results are  affected by \emph{external validity}, where different conclusions can be drawn on compliance-readiness of datasets in other locations (e.g. Zenodo) or for datasets that use a schema for which column names could not be retrieved. Since many dataset cards use external references (e.g. publications), information that was not available on HF but could be found on a research paper or a GitHub repository may have been missed. Similarly, if information for data preparation and data used is provided in a different location in the dataset card and not in the main sections where this information is usually expected, the respective information could not have been considered. This may have affected \emph{conclusion validity}. Nevertheless, our aim was to examine HF alone and the extent to which information on compliance-readiness is available within HF. 

We also did not focus on locating facial recognition and voice data that are a special case of compliance (e.g. as in the Illinois law that concerns facial recognition).
In order to ensure that we calculated only cases with valid content (and not acting only as a placeholder), we have performed a manual analysis to detect cases where information from HF or relevant documentation was added, instead of a valid section content. We then filtered out those cases, in order to address a relevant \emph{construct validity} threat arising from this observation.

\textbf{Future implications.} These preliminary results are an indication that standards and tools are required for compliance identification and automation in HF datasets, with specific implications for the following target groups:
\begin{itemize}
    \item \emph{Researchers}: More research should be directed towards regulatory compliance processes. Coding tools that rely on AI agents (e.g. Codex) may be used in HF in order to ensure proper inclusion of sections in dataset cards. Moreover, we found cases of datasets that refer to other HF datasets in their data sources, so dataset reuse is an area that requires further investigation.
    \item \emph{Practitioners}: Dataset cards need to ensure that all relevant information is included, and existing tools can be used before uploading a dataset to ensure no PII or sensitive data are leaked or used in models' training. Overall, our results show that the use of scrubbing techniques is rare. Since automated tools exist for this process and our analysis showed that they are used in a limited degree (e.g. Presidio), a good guideline for compliance is that it is useful to use them in order to automate the process.
    \item \emph{Platforms and Regulators}: Dataset cards analysis showed that there are scarce mentions to data cleaning and preparation processes. More guidance is needed on how AI datasets need to be documented for compliance but also transparency purposes. Detailed platform guidelines might also help towards this direction.
\end{itemize}

\section{Conclusions}

In this paper, we presented a preliminary work on readiness of AI datasets for regulatory compliance. The work focused on metadata and keyword heuristics from HF dataset cards and data schemas to examine whether practices towards regulatory compliance (e.g. details of data processing techniques) are present. We have also investigated to which degree dataset schemas include mentions to PII and other sensitive data. The results show that most datasets do not pay special attention to such properties, as the respective sections are missing in most cases. Nevertheless, some datasets might potentially expose user data and it should be investigated whether user consent was given. Our future work will explore the intersection of HF dataset cards and GitHub information for datasets by analyzing in more detail the available information (e.g. outside HF). We will also work on the creation of tools that automate the process of providing more information for compliance purposes, as this was identified as a direct need from the current analysis.

\bibliographystyle{splncs04}
\bibliography{sample-base}

\end{document}